\documentstyle[12pt,epsfig]{report}

\newcommand{\nonumsection}[1] {\vspace{12pt}\noindent{\bf #1}
        \par\vspace{5pt}}

\def\thebibliography#1{\nonumsection{\large \it References}\list
  {[\arabic{enumi}]}{\settowidth\labelwidth{[#1]}\leftmargin\labelwidth
    \advance\leftmargin\labelsep
    \usecounter{enumi}}
    \def\newblock{\hskip .11em plus .33em minus .07em}
    \sloppy\clubpenalty4000\widowpenalty4000}

\newcommand{\smalllineskip}{\baselineskip=12pt}

\newcommand{\be}{\begin{eqnarray}}
\newcommand{\ee}{\end{eqnarray}}

\begin{document}

\baselineskip=18pt

\rightline{UNITU-THEP-7/1998}
\rightline{April (revised May) 1998}
\rightline{hep-ph/9804260}

\vskip 1.5truecm
\centerline{\Large\bf Radial excitations of low--lying baryons}

\bigskip

\centerline{\Large\bf and the $Z^+$ penta--quark}
\vskip1.0cm
\centerline{Herbert Weigel}
\vskip .5cm
\centerline{Institute for Theoretical Physics}
\centerline{T\"ubingen University}
\centerline{Auf der Morgenstelle 14}
\centerline{D-72076 T\"ubingen, Germany}
\vskip 1.5cm

\centerline{\bf ABSTRACT}

\vskip 0.5cm
\parbox[t]{14cm}{
\baselineskip=20pt
Within an extended Skyrme soliton model for baryons the interplay 
between the collective radial motion and the SU(3)--flavor--rotations 
is investigated. The coupling between these modes is mediated by 
flavor symmetry breaking. Collective coordinates which describe the 
corresponding large amplitude fluctuations are introduced and treated 
canonically. When diagonalizing the resulting Hamiltonian flavor 
symmetry breaking is fully taken into consideration. As eigenstates 
not only the low--lying $\frac{1}{2}^+$ and $\frac{3}{2}^+$ baryons but 
also their radial excitations are obtained and compared to the empirical 
data. In particular the relevance of radial excitations for the 
penta--quark baryon $Z^+$ ($Y=2$, $I=0$, $J^\pi=\frac{1}{2}^+$) is 
discussed. In this approach its mass is predicted to be $1.58{\rm GeV}$. 
Furthermore the widths for various hadronic decays are estimated which, 
for example, yields $\Gamma(Z^+\to NK)\sim 100{\rm MeV}$ for the only 
permissible decay process of the $Z^+$.}

\vskip 1.5cm
\noindent
\leftline{\it PACS: 12.39.Dc, 14.20.Gk, 14.20.Jn.}
\vfill\eject

\baselineskip=20pt
\stepcounter{chapter}
\leftline{\large\it 1. Introduction}
\smallskip

Recently there has been renewed interest in baryon states which cannot be 
described as simple bound states of three quarks \cite{Wa92,Wa94,Di97}. 
One of the most prominent examples is the so--called $Z^+$ which possesses 
the spin and isospin quantum numbers of the $\Lambda$ hyperon, however, it 
carries hypercharge $Y=2$. When extending chiral soliton models to flavor 
SU(3) \cite{Gu84,We96} such states come about quite naturally as they are 
members of higher dimensional representations which do not have 
counterparts of equal quantum numbers in the octet or decuplet. 

In the context of chiral soliton models these higher dimensional 
representations have gained most of their recognition from the 
investigation of flavor symmetry breaking. Besides such {\it exotic} 
states as the $Z^+$, the higher dimensional representations also contain 
states which have the spin and flavor quantum numbers of the low--lying 
$\frac{1}{2}^+$ and $\frac{3}{2}^+$ baryons. One easily recognizes that 
flavor symmetry breaking couples states which belong to different SU(3) 
representations but otherwise have identical quantum numbers. Hence these 
states mix with the octet and decuplet baryons as the higher dimensional 
representations provide a basis to obtain the exact eigenstates of the 
full collective Hamiltonian \cite{Ya88}. As a consequence the nucleon is 
no longer a pure octet state but also contains sizable admixture of the 
corresponding members in the ${\overline{\bf 10}}$ and $\bf 27$ dimensional 
representations \cite{Pa89,Lee89}. Permissible representations are those which
contain a non--strange baryon state with identical spin and isospin 
\cite{Ma84}.

Presumably the lightest state which does not have a counterpart of 
equal quantum numbers in the octet or decuplet is the above mentioned 
$Z^+$. The lowest dimensional representation containing a state with 
the quantum numbers of the $Z^+$ is the ${\overline{\bf 10}}$. When 
glancing at the Young tableau of the ${\overline{\bf 10}}$,
\raisebox{-4pt}{\epsfxsize=1cm\epsfysize=0.5cm\epsffile{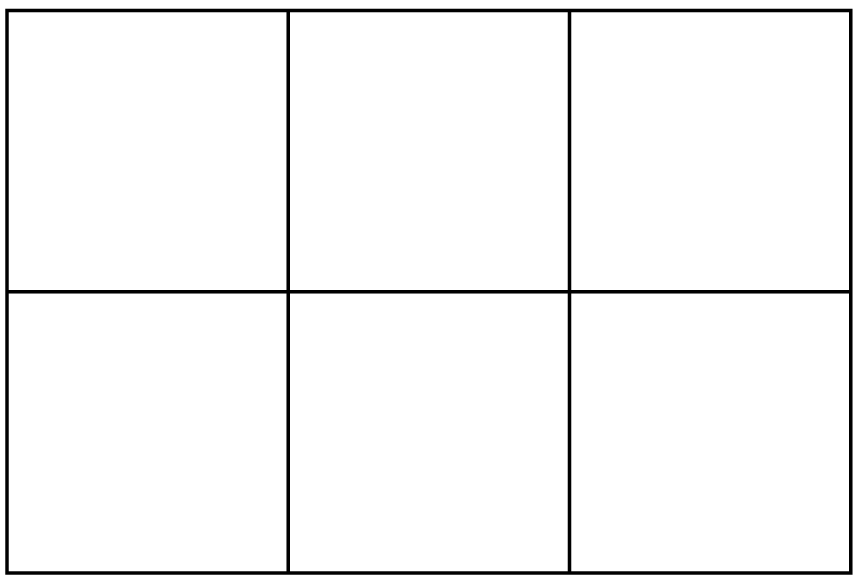}},
it becomes immediately clear that such states are not simple bound states 
containing only three quarks. Rather they have to be interpreted as
a quark--antiquark pair coupled to a three quark state. Such objects
are commonly called penta--quarks. In refs \cite{Wa92,Wa94} quantitative
calculations for the mass of the $Z^+$ were performed within the Skyrme 
model \cite{Sk61,Ad83,Ho86}. From fig. 2 of ref \cite{Wa92} one deduces
that the $Z^+$ should be about $0.7{\rm GeV}$ heavier than the nucleon. 
This is not too different from the estimate of ref \cite{Di97} where a mass 
difference of $0.59{\rm GeV}$ with respect to the nucleon was predicted. 
However, the latter prediction is just $100{\rm MeV}$ above the 
threshold for the only accessible decay process, $Z^+\to N K$. While
the result obtained in ref \cite{Wa92} stems from a self--contained 
model calculation the authors of ref \cite{Di97} collected 
almost\footnote{For example, the admixture of the states of the 
$\bf 27$--plet to the octet as well as the coupling between the 
$Z^+$--type states in the ${\overline{\bf 10}}$ and ${\overline{\bf 35}}$ 
representations were omitted.} all contributions up to linear order in 
flavor symmetry breaking to the collective Hamiltonian and the baryon 
wave--functions which are consistent with the general transformation 
properties of these objects in flavor space. The associated constants of 
proportionality were determined from data for the established baryons. In 
particular (up to flavor symmetry breaking effects) the $N(1710)$ was 
identified with the nucleon state in the ${\overline{\bf 10}}$ representation 
in order to fix the mass difference of the states within that representation 
to the octet baryons. This treatment is not without ambiguities because not 
all baryons are rotational excitations. In such a picture one first wonders 
about the role of the Roper (1440) resonance which in Skyrmion models often 
is identified as the radial (or breathing) $\hbar\omega$ excitation of the 
nucleon \cite{Ha84,HH84,Ab95}. Secondly it is then natural to consider the 
$N(1710)$ resonance as the corresponding $2\hbar\omega$ excitation. This 
scenario leads to the obvious question whether there is an interplay between 
radial and rotational excitations and especially to what extent this 
interplay effects the predictions for the $Z^+$. The coupling between 
these two types of excitations is mediated by flavor symmetry breaking. 
For the ordinary baryons this interplay has already been discussed some 
time ago \cite{Sch91,Sch91a}. It is the purpose of the present study to 
extend this approach to the $Z^+$ for which the radial motion has not been
considered previously.

In section 2 the simultaneous treatment of collective coordinates 
for radial and rotational motion of the soliton will be discussed for 
a special chiral model. Also the resulting baryon spectrum will be
compared to the empirical data. In section 3 the widths for various 
hadronic decay modes of excited baryons will be estimated. Section 4 
serves to summarize the present study.

\bigskip
\stepcounter{chapter}
\leftline{\large\it 2. Breathing mode approach in flavor SU(3)}
\smallskip

In this section we will describe the treatment of large amplitude 
fluctuations for radial and rotational degrees of freedom in a three 
flavor soliton model. The simplest model within which such a study 
can be carried out is the Skyrme model with only pseudoscalar fields. 
Unfortunately, the breathing mode approach to this model does not 
adequately reproduce the mass differences in baryon spectrum \cite{Sch91}. 
For the present investigation we will therefore employ an extended 
version of the Skyrme model by supplementing it with a scalar 
field as motivated by the trace anomaly of QCD. Although the breathing 
mode approach to this model reasonably describes the baryon spectrum as 
well as various static baryon properties \cite{Sch91a} it seems somewhat 
unmotivated. A more natural model choice would rather employ vector meson 
\cite{Me88} or chiral quark \cite{Al96} models. The reason being that 
these mesonic degrees of freedom need to be included in order to obtain 
non--vanishing neutron--proton mass differences as well as a finite axial 
singlet current matrix element of the nucleon \cite{Ja89}. In these models 
meson fields, which vanish classically, are induced by the collective 
rotation $A(t)$ in eq (\ref{colcoor}). As it is yet unknown how to treat 
the breathing mode in the presence of these induced fields the model 
may be considered as an effective parameterization of massive meson 
fields. A particular difficulty with these induced fields is that the 
corresponding stationary conditions must separately be solved for every 
value of the scaling coordinate $x$ in order to maintain the correct 
normalization of the Noether currents. In vector meson models this 
problem occurs already on the classical level for the time component of 
the $\omega$ field because its stationary condition actually is a 
constraint which guarantees the positivity of the classical energy 
functional. The incorporation of the breathing coordinate in an SU(2) 
vector meson model was attempted in ref \cite{Ma90}, however, the above 
mentioned subtleties were ignored. In non--topological chiral quark 
soliton models their non--confining character may cause additional 
problems when treating the breathing mode dynamically because a 
transition to the trivial meson configuration may occur \cite{Ab95}.

To be specific we will follow the treatment of ref
\cite{Sch91a} where the soliton model not only contains the pseudoscalar 
mesons $\phi^a$ but also an effective scalar meson field 
$H=\langle H \rangle {\rm exp}(4\sigma)$ which is introduced to mock up the 
QCD anomaly \cite{Go86} for the dilatation current 
\be
-\partial_\mu D^\mu= H + \sum_i m_i{\bar \Psi}_i\Psi_i
\quad {\rm where} \quad 
H=-\frac{\beta(g)}{g} G_{\mu\nu}^a G^{a\mu\nu} .
\label{dilano}
\ee
The vacuum expectation value $\langle H \rangle\sim
(0.30-0.35 {\rm GeV})^4$ can be extracted from sum rule estimates for the
gluon condensate \cite{Sh79}. Eventually the fluctuating field $\sigma$
may be identified as a scalar glueball. The effective mesonic action reads
\be
\Gamma=\int d^4x \left({\cal L}_0+{\cal L}_{\rm SB}\right)
+\Gamma_{\rm WZ} \ .
\label{act1}
\ee
The flavor symmetric part involves both the chiral field\footnote{Here 
the normalization coefficients $f_a$ refer to the pseudoscalar decay 
constants.} $U={\rm exp} (i\lambda^a\phi^a/f_a)$ 
as well as the scalar gluonic fluctuation $\sigma$
\be
{\cal L}_0&=&-\frac{f_\pi^2}{4}{\rm e}^{2\sigma}
{\rm tr}\left(\alpha_\mu\alpha^\mu\right)
+\frac{1}{32e^2}{\rm tr}
\left(\left[\alpha_\mu,\alpha_\nu\right]^2\right)
\nonumber \\
&&+\frac{1}{2}\Gamma_0^2\ {\rm e}^{2\sigma} \ 
\partial_\mu\sigma\partial^\mu\sigma
+{\rm e}^{4\sigma}\left\{\frac{1}{4}
\left[\langle H \rangle 
-6\left(2\delta^\prime+\delta^{\prime\prime}\right)\right]
- \sigma\langle H \rangle\right\}
\label{act2}
\ee
with $\alpha_\mu=\partial_\mu U U^\dagger$. Assuming the canonical 
dimensions $d(U)=0$ and $d(H)=4$ it is straightforward to verify 
that (\ref{act2}) yields the anomaly equation (\ref{dilano}) for 
$m_i=0$. The terms which lift the degeneracy between mesons of different 
strangeness are comprised in 
\be
{\cal L}_{\rm SB}={\rm tr}\Big\{
\left(\beta^\prime{\hat T}+\beta^{\prime\prime}{\hat S}\right)
{\rm e}^{2\sigma}\partial_\mu U \partial^\mu U^\dagger U
+\left(\delta^\prime{\hat T}+\delta^{\prime\prime}{\hat S}\right)
{\rm e}^{3\sigma}U+{\rm h. c.}\Big\}\ ,
\label{act3}
\ee
where the flavor projectors ${\hat T}={\rm diag}(1,1,0)$ and 
${\hat S}={\rm diag}(0,0,1)$ have been introduced. Using a sigma--model 
interpretation of the chiral field the coupling of the scalar field in 
${\cal L}_{\rm SB}$ is such as to reproduce the explicit breaking in 
the anomaly equation (\ref{dilano}) \cite{Ad87}. The major impact 
of the scalar field emerges through the factor ${\rm e}^{3\sigma}$ in 
the mass term of the symmetry breaking piece (\ref{act3}). As will be 
discussed later, this mitigates the symmetry breaking effects in the 
baryon sector. This factor is special to the model with the trace anomaly 
included since it properly accounts for the explicit breaking of the 
dilatation current (\ref{dilano}) as the quark bilinear 
${\bar \Psi}_i\Psi_i$ has canonical mass dimension three.

The various parameters 
in eqs (\ref{act2}) and (\ref{act3}) are determined from the masses 
and decay constants of the pseudoscalar mesons:
\be
\beta^\prime\approx 26.4 {\rm MeV}^2,\
\beta^{\prime\prime}\approx 985 {\rm MeV}^2,\
\delta^\prime\approx 4.15\times10^{-5} {\rm GeV}^2,\
\delta^{\prime\prime}\approx 1.55\times10^{-3}{\rm GeV}^4\ .
\label{para1}
\ee
Then the only free parameters of the model are the Skyrme 
constant $e$ and the glueball mass,
\be
m_\sigma^2=\frac{4\langle H \rangle +
6(2\delta^\prime+\delta^{\prime\prime})}{\Gamma_0^2} \ .
\label{msig}
\ee
As in ref \cite{Sch91a} we will use $m_\sigma\approx1.25{\rm GeV}$.
Finally the scale invariant Wess--Zumino term \cite{WZW} is most 
conveniently presented by introducing the one--form 
$\alpha=\alpha_\mu dx^\mu$,
\be
\Gamma_{\rm WZ}=\frac{iN_c}{240\pi^2}\int {\rm tr}(\alpha^5) \ .
\label{wzterm}
\ee
The above described model possesses a static soliton solution 
$U_0(\mbox{\boldmath $r$})={\rm exp}
[i \mbox{\boldmath $\tau$}\cdot \hat{\mbox{\boldmath $r$}}F(r)]$,
$\sigma(\mbox{\boldmath $r$})=\sigma_0(r)$ which is characterized
by the two radial functions $F(r)$ and $\sigma_0(r)$ \cite{Go86,Ja87}. 
Except of unit baryon number this configuration does not carry baryonic
quantum numbers such as spin or isospin. Baryon states are commonly 
generated by canonical quantization of the collective coordinates which 
are introduced to describe large amplitude fluctuations. Apparently these 
are the rotations in coordinate and flavor spaces which are (up to flavor 
symmetry breaking) zero modes of the soliton. Due to the hedgehog 
structure of the soliton these rotations are equivalent. In addition the 
energy surface associated with scale or breathing transformations of the 
Skyrmion is known to be flat, at least in a large vicinity of the 
stationary point \cite{Ha84,HH84}. For this reason it is suggestive to 
also introduce a collective coordinate for the soliton extension. Then 
the unknown time--dependent solution to the Euler equations is 
approximated by
\be
U(\mbox{\boldmath $r$},t) = 
A(t) U_0\left(\mu(t)\mbox{\boldmath $r$}\right) A^\dagger(t)
\quad {\rm and}\quad
\sigma(\mbox{\boldmath $r$},t)=\sigma_0\left(\mu(t)r\right) \ .
\label{colcoor}
\ee
Substituting this parameterization  into the action (\ref{act1})
yields the Lagrangian for the collective coordinates $A(t)$ as well as
$x(t)=[\mu(t)]^{-3/2}$
\be
L(x,\dot x,A,\dot A)\hspace{-0.1cm}&=&\hspace{-0.1cm}{4\over9}
\left(a_1+a_2x^{-{4\over3}}\right){\dot x^2}-
\left(b_1x^{2\over3}+b_2x^{-{2\over3}}+b_3x^2\right)
+{1\over2}\left(\alpha_1x^2+\alpha_2x^{2\over3}\right)
\sum^3_{a=1}\Omega^2_a
\nonumber \\* && \hspace{-1.5cm}
+{1\over2}\left(\beta_1x^2+\beta_2x^{2\over3}\right)
\sum^7_{a=4}\Omega^2_a +{\sqrt3\over2}\Omega_8
-\left(s_1x^2+s_2x^{2\over3}+{4\over9}s_3{\dot x^2}\right)
\left(1-D_{88}\right) \ .
\label{lagbreath}
\ee
Here the angular velocities 
$A^\dagger \dot A = (i/2)\sum_{a=1}^8 \lambda_a\Omega_a$ as well as 
the adjoint representation 
$D_{ab}=(1/2){\rm tr}(\lambda_a A \lambda_b A^\dagger)$ have been 
introduced. A term linear in $\dot x$, which would originate from 
flavor symmetry breaking terms, has been omitted because the matrix 
elements of the associated $SU(3)$ operators vanishes when properly 
accounting for Hermiticity in the process of quantization \cite{Pa91}. 
The expressions for the constants $a_1,\ldots,s_3$ as functionals of
the chiral angle as well as their numerical values may be extracted
from ref \cite{Sch91a}. The term involving $s_3$ causes major 
difficulties in the process of quantization. This contribution to $L$ 
stems from the derivative type symmetry breaker in (\ref{act3}) whose 
influence in the soliton sector is known to be small\footnote{In 
numerical calculations the direct contributions of this term are small. 
Nevertheless it is important because it has significant indirect 
influence since it provides the origin for different decay 
constants, $f_\pi\approx 1.2f_K$. Compared to the unphysical case 
$f_\pi=f_K$ the mass type symmetry breaker increases by about 50\% 
because $\delta^{\prime\prime}=(2f_K^2m_K^2-f_\pi^2m_\pi^2)/4$. The 
$\delta^{\prime\prime}$ term is contained in $s_1$.}. In addition, 
by replacing the collective function $1-D_{88}$ with a constant 
of order unity the effects of this term have been estimated to be 
only a few percent, {\it cf.} appendix B of ref \cite{Sch91}. Hence 
the $s_3$ term may safely be omitted.

The baryon states corresponding to the Lagrangian (\ref{lagbreath}) are 
obtained in a two--step procedure. In the first step flavor symmetry 
breaking is ignored. For convenience one furthermore defines
\be
m&=&m(x)=\frac{8}{9}(a_1+a_2x^{-{4\over3}})\ , \quad
b=b(x)=b_1x^{2\over3}+b_2x^{-{2\over3}}+b_3x^2\ , 
\nonumber \\
\alpha&=&\alpha(x)=\alpha_1x^2+\alpha_2 x^{\frac{2}{3}}\ , \quad
\beta=\beta(x)=\beta_1x^2+\beta_2 x^{\frac{2}{3}}
\nonumber 
\ee
and
\be
s&=&s(x)=s_1x^2+s_2x^{\frac{2}{3}} \ .
\label{defbreath}
\ee
Then the flavor symmetric part of the collective Hamiltonian
\be
H=-\frac{1}{2\sqrt{m\alpha^3\beta^4}}\frac{\partial}{\partial x}
\sqrt{\frac{\alpha^3\beta^4}{m}}\frac{\partial}{\partial x}
+b+\left(\frac{1}{2\alpha}-\frac{1}{2\beta}\right)J(J+1)
+\frac{1}{2\beta}C_2(\mu)-\frac{3}{8\beta}+s 
\label{freebreath}
\ee
is diagonalized for a definite $SU(3)$ representation $\mu$. Due to the 
hedgehog structure of the static configuration $U_0$ and $\sigma_0$, the 
allowed representations must contain at least one state with identical 
spin and isospin. In addition, this state must have vanishing strangeness 
\cite{Gu84,We96}. For definiteness we denote
the eigenvalues of (\ref{freebreath}) by ${\cal E}_{\mu,n_\mu}$ and 
the corresponding eigenstates by $|\mu,n_\mu\rangle$, where $n_\mu$ 
labels the radial excitations. Actually the eigenstates factorize
$|\mu,n_\mu\rangle=|\mu\rangle|n_\mu\rangle$. In this language the
nucleon corresponds to $|{\bf 8},1\rangle$ while the first radially 
excited state, which is commonly identified with the Roper (1440) 
resonance, would be $|{\bf 8},2\rangle$. Of course, we are interested 
in the role of states like
$|{\overline {\bf 10}},n_{\overline {\bf 10}}\rangle$ since in particular 
this tower contains the state with the quantum numbers of the $Z^+$. In 
the second step the symmetry breaking part will be taken into account. This 
is done by employing the states $|\mu,n_\mu\rangle$ as a basis to 
diagonalize the complete Hamiltonian matrix
\be
H_{\mu,n_\mu;\mu^\prime,n^\prime_{\mu^\prime}}=
{\cal E}_{\mu,n_\mu}\delta_{\mu,\mu^\prime}
\delta_{n_\mu,n^\prime_{\mu^\prime}}
-\langle\mu|D_{88}|\mu^\prime\rangle
\langle n_\mu|s(x)|n^\prime_{\mu^\prime}\rangle \ .
\label{hammatr}
\ee
The flavor part of these matrix elements is computed using $SU(3)$ 
Clebsch--Gordon coefficients\footnote{The Clebsch--Gordon coefficients 
not provided in ref \cite{Sw63} are numerically computed as described in 
footnote 14 of ref \cite{Sch91} based on the Euler angle decomposition
of ref \cite{Ya88}.} while the radial part is calculated numerically
using the appropriate eigenstates of (\ref{freebreath}). Of course, this 
can be done for each isospin multiplet separately, {\it i.e.} flavor 
quantum numbers are not mixed. The physical baryon states $|B,m\rangle$ are 
finally expressed as linear combinations of the eigenstates of
the symmetric part
\be
|B,m\rangle=\sum_{\mu,n_\mu}C_{\mu,n_\mu}^{(B,m)}
|\mu,n_\mu\rangle \ .
\label{bsbreath}
\ee
The corresponding eigenenergies are denoted by $E_{B,m}$.
The nucleon $|N,1\rangle$ is then identified as the lowest energy
solution with the associated quantum numbers, while the Roper
is defined as the next state ($|N,2\rangle$) in the same spin --
isospin channel. Turning to the quantum numbers of the $\Lambda$
provides not only the energy $E_{\Lambda,1}$ and
wave--function $|\Lambda,1\rangle$ of this hyperon but also the
analogous quantities for the radially excited $\Lambda$'s:
$E_{\Lambda,m}$ and $|\Lambda,m\rangle$ with $m\ge2$. These
calculations are repeated for the other spin -- isospin channels
yielding the spectrum not only of the ground state $\frac{1}{2}^+$
and $\frac{3}{2}^+$ baryons but also their radial excitations. 
Of course, flavor symmetry breaking couples all possible 
$SU(3)$ representations. When diagonalizing (\ref{hammatr})
we consider the basis built by the representations 
$\mbox{\boldmath $8$}$, $\overline{\mbox{\boldmath $10$}}$,
$\mbox{\boldmath $27$}$, $\overline{\mbox{\boldmath $35$}}$,
$\mbox{\boldmath $64$}$, $\overline{\mbox{\boldmath $81$}}$,
$\mbox{\boldmath $125$}$, $\overline{\mbox{\boldmath $154$}}$ for
the $\frac{1}{2}^+$ baryons and
$\mbox{\boldmath $10$}$, ${\mbox{\boldmath $27$}}$,
$\mbox{\boldmath $35$}$, $\overline{\mbox{\boldmath $35$}}$,
$\mbox{\boldmath $64$}$, $\overline{\mbox{\boldmath $28$}}$,
$\mbox{\boldmath $81$}$, $\overline{\mbox{\boldmath $81$}}$
$\mbox{\boldmath $125$}$, $\overline{\mbox{\boldmath $80$}}$
$\mbox{\boldmath $154$}$, $\overline{\mbox{\boldmath $254$}}$ for
the $\frac{3}{2}^+$ baryons. For the breathing degree of freedom we
include basis states which are up to $4{\rm GeV}$ above the ground
states of the flavor symmetric piece (\ref{freebreath}), {\it i.e.} 
$|\mbox{\boldmath $8$},1\rangle$ and $|\mbox{\boldmath $10$},1\rangle$ 
for the $\frac{1}{2}^+$ and $\frac{3}{2}^+$ baryons, respectively. This 
seems to be sufficient to get acceptable convergence when diagonalizing 
(\ref{hammatr}). It should be noted that not all of the above SU(3) 
representations appear in each isospin channel. For example, there is no 
$\Lambda$--type state in the $\overline{\mbox{\boldmath $10$}}$. 

In table \ref{tab_1} the predictions 
for the mass differences\footnote{In soliton models commonly only 
mass differences are considered to avoid the inclusion of meson loop 
corrections which reduce the absolute values substantially \cite{Mo93}.} 
with respect to the nucleon of the eigenstates are shown for two values 
of the Skyrme parameter $e$.
\begin{table}
\caption{\label{tab_1}
The mass differences with respect to the nucleon (939MeV) of the 
eigenstates of the Hamiltonian (\protect\ref{hammatr}). Experimental data 
are taken from \protect\cite{PDG}, if available. It should be remarked 
that even single--star resonances have been included. The notation for 
the states appearing in this table is defined in
eq (\protect\ref{bsbreath}). All numbers are in MeV.}
~
\newline
\centerline{\tenrm\smalllineskip
\begin{tabular}{c | c c c | c c c | c c c}
B& \multicolumn{3}{c|}{$m=0$} & \multicolumn{3}{c|}{$m=1$}
& \multicolumn{3}{c}{$m=2$} \\
\hline
& $e$=5.0 & $e$=5.5 & expt. 
& $e$=5.0 & $e$=5.5 & expt.
& $e$=5.0 & $e$=5.5 & expt. \\ 
\hline
N & \multicolumn{3}{c|}{Input}
            & 413 & 445 & 501 & 836 & 869 & 771  \\
$\Lambda$ 
& 175 & 173 & 177 & 657 & 688 & 661 & 1081 & 1129 & 871 \\
$\Sigma$
& 284 & 284 & 254 & 694 & 722 & 721 & 1068 & 1096 & 838 \\
$\Xi$
& 382 & 380 & 379 & 941 & 971 & --- & 1515 & 1324 & --- \\
\hline
$\Delta$
& 258 & 276 & 293 & 640 & 680 & 661 & 974 & 1010 & 981 \\
$\Sigma^*$
& 445 & 460 & 446 & 841 & 878 & 901 & 1112 & 1148 & 1141 \\
$\Xi^*$
& 604 & 617 & 591 & 1036 & 1068 & --- & 1232 & 1269 & --- \\
$\Omega$
& 730 & 745 & 733 & 1343 & 1386 & --- & 1663 & 1719 & --- \\
\end{tabular}}
\end{table}
The agreement with the experimental data is quite astonishing, not
only for the ground state but also for the radial excitations. Only 
the prediction for the Roper resonance ($|N,2\rangle$) is on the low 
side. This is common to the breathing mode approach \cite{Ha84,HH84}.
As far as data are available the other first excited states are 
quite well reproduced. On the other hand for the $\frac{1}{2}^+$ baryons 
the energy eigenvalues for the second excitations overestimate the 
corresponding empirical data somewhat. However, the pattern 
$M(|N,2\rangle)<M(|\Sigma,2\rangle)<M(|\Lambda,2\rangle)$ is reproduced. 
The predicted $\Sigma$ and $\Lambda$ type states with $m=2$ are about 
$200{\rm MeV}$ too high. For the $\frac{3}{2}^+$ baryons with $m=2$ the 
agreement with data is much better, on the 3\% level. On the whole, the 
present model gives fair agreement with the available data. This certainly 
supports the picture of coupled radial and rotational modes.

Above we stated that the factor ${\rm e}^{3\sigma}$ in eq (\ref{act3}) 
mitigated the effects of flavor symmetry breaking in the soliton sector. 
For the soliton solution the $\sigma$ field is always negative. Hence
the contribution of the mass--type symmetry breaker to $s(x)$ in 
eq (\ref{defbreath}) is significantly reduced as compared to the pure 
pseudoscalar case. As already discussed, the mass--type symmetry breaker 
strongly dominates over the kinetic--type, which is suppressed by the 
factor ${\rm e}^{2\sigma}$. Since applying the breathing mode
approach to the pure pseudoscalar model ({\it i.e.} $\sigma\equiv0$)
overestimates the flavor symmetry breaking in the baryon mass differences
\cite{Sch91}, the incorporation of the scalar glueball field 
improves on the predicted baryon spectrum \cite{Sch91a}. 
In the pure Skyrme model a reduction of symmetry breaking effects can 
be gained by decreasing the Skyrme constant $e$. Unfortunately this 
also lowers the difference between the nucleon and the $\Delta$ masses. 
As can be observed from table 1 in ref \cite{Sch91} an overall satisfactory 
picture cannot be obtained in the breathing mode approach to the pure 
pseudoscalar model.

\begin{figure}[ht]
\centerline{\hskip -1.0cm
\epsfig{figure=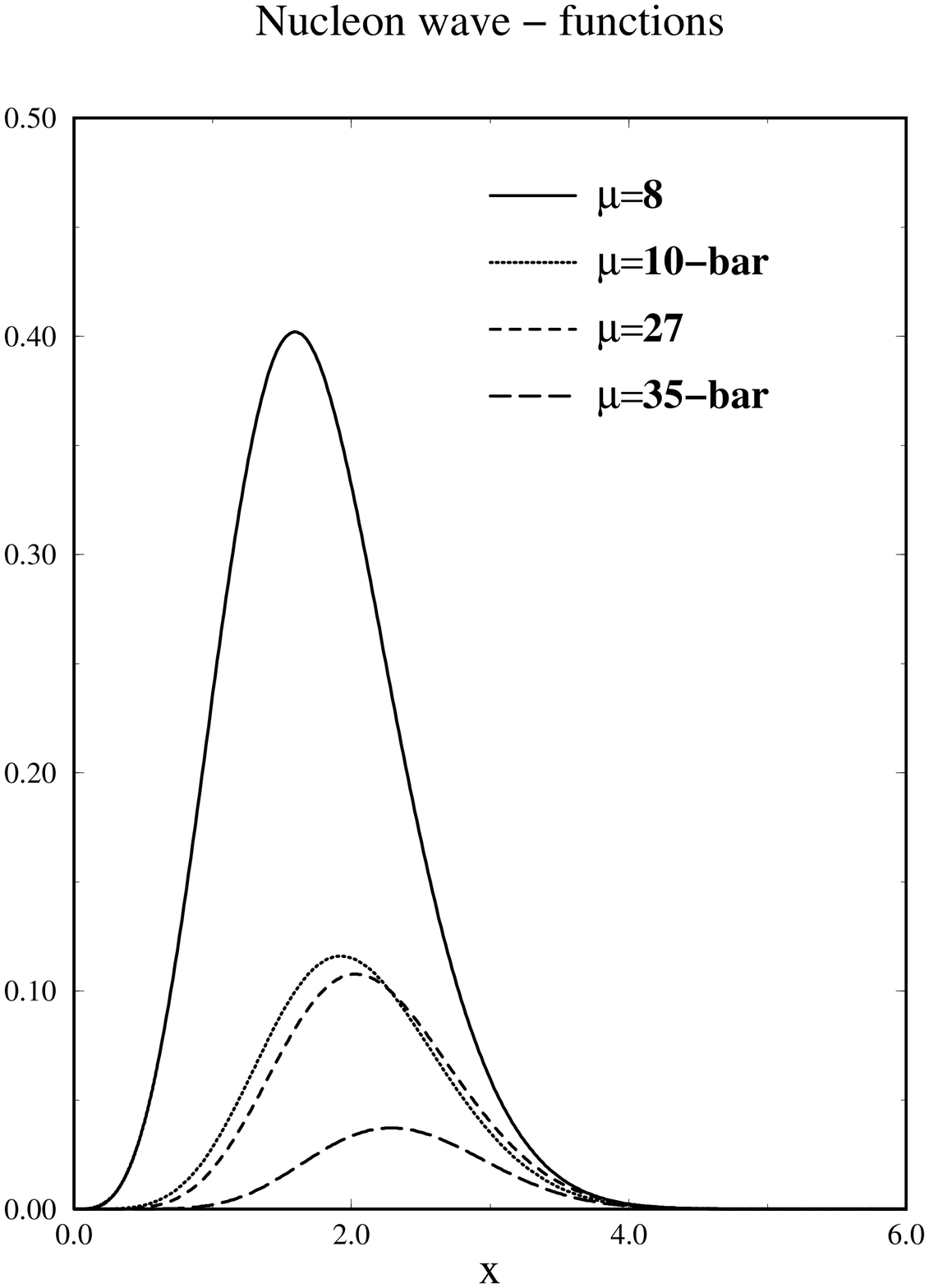,height=6cm,width=7cm}
\epsfig{figure=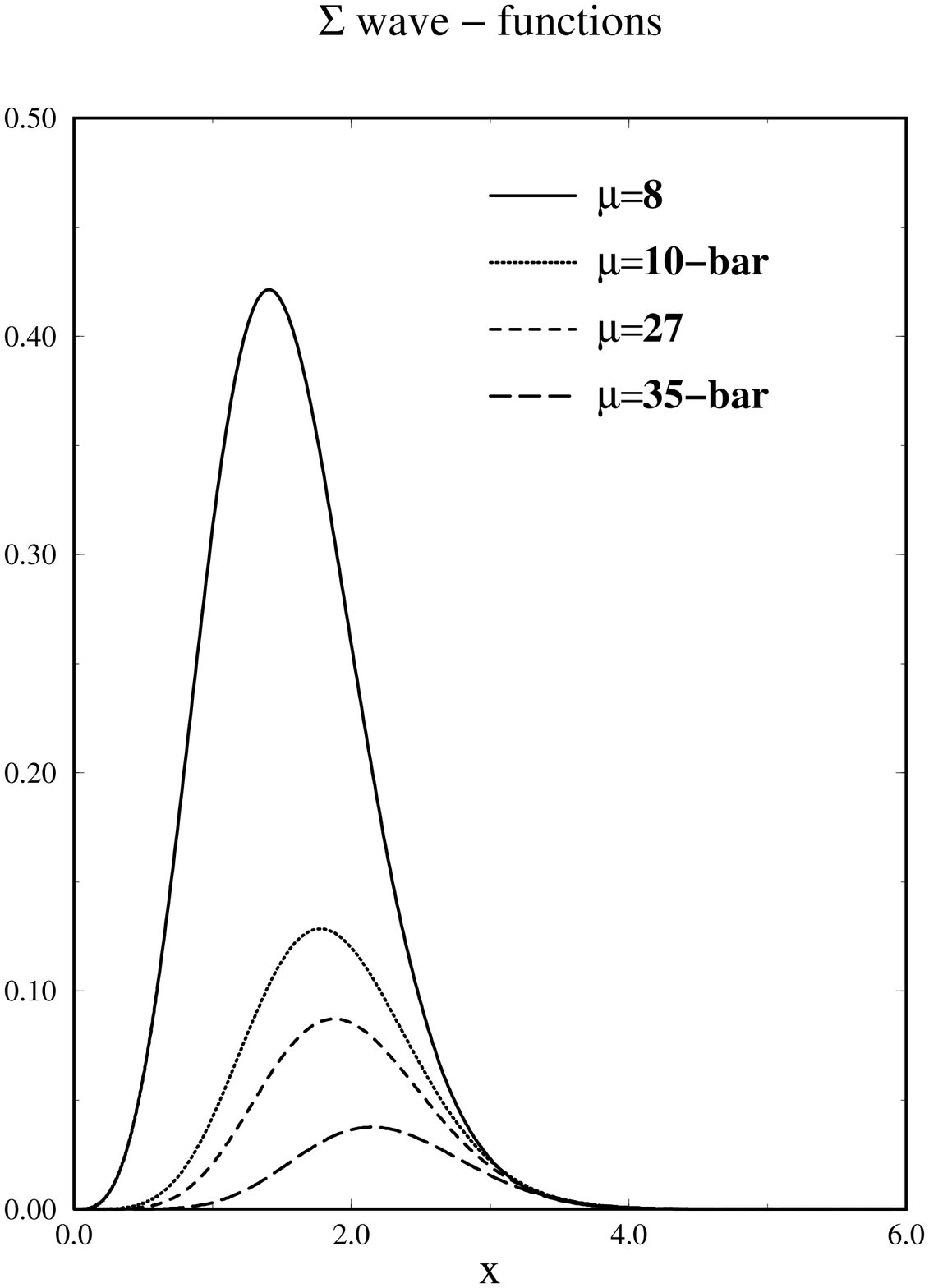,height=6cm,width=7cm}}
\vskip1.8cm
\centerline{\hskip -1.0cm
\epsfig{figure=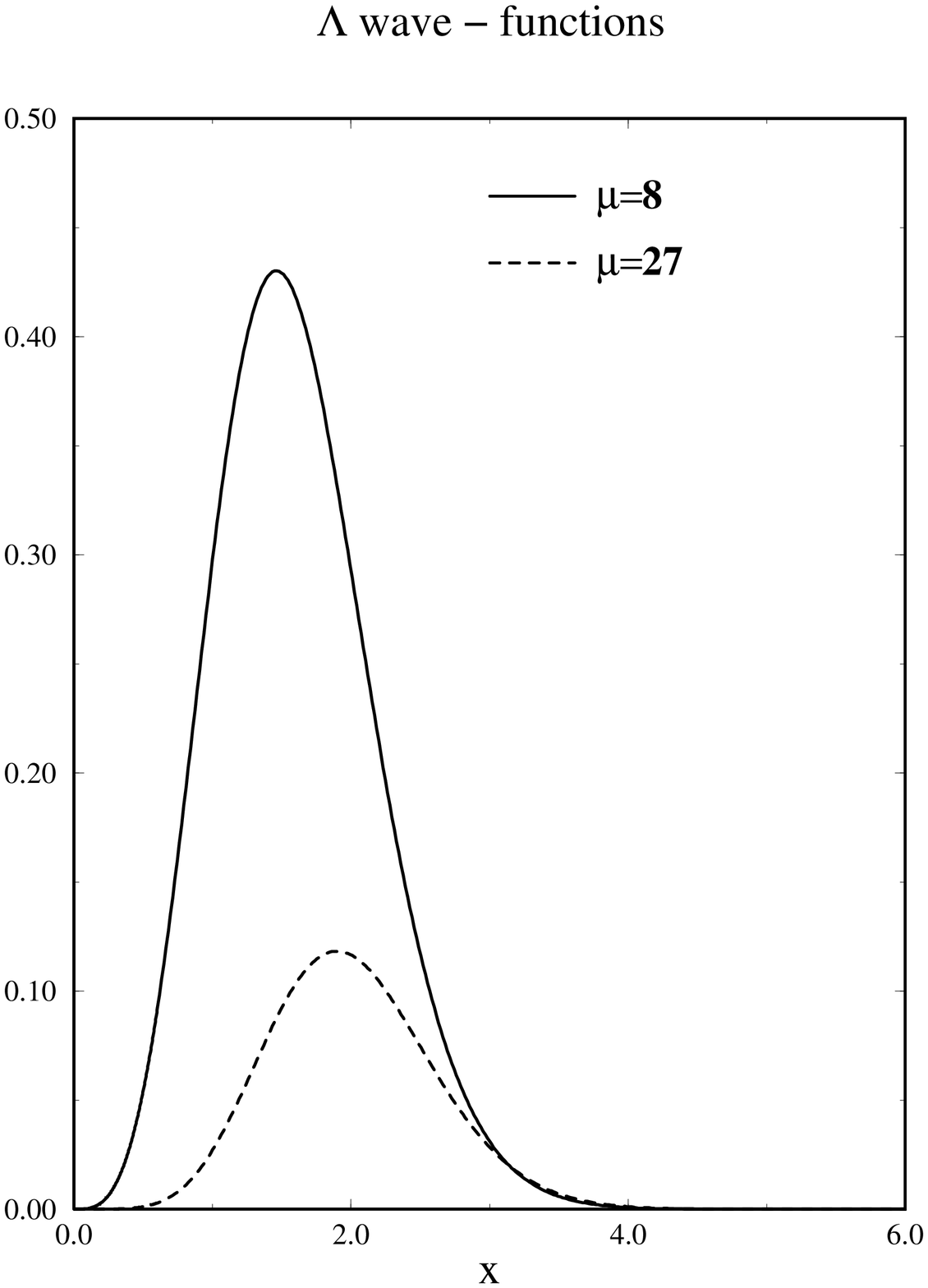,height=6cm,width=7cm}
\epsfig{figure=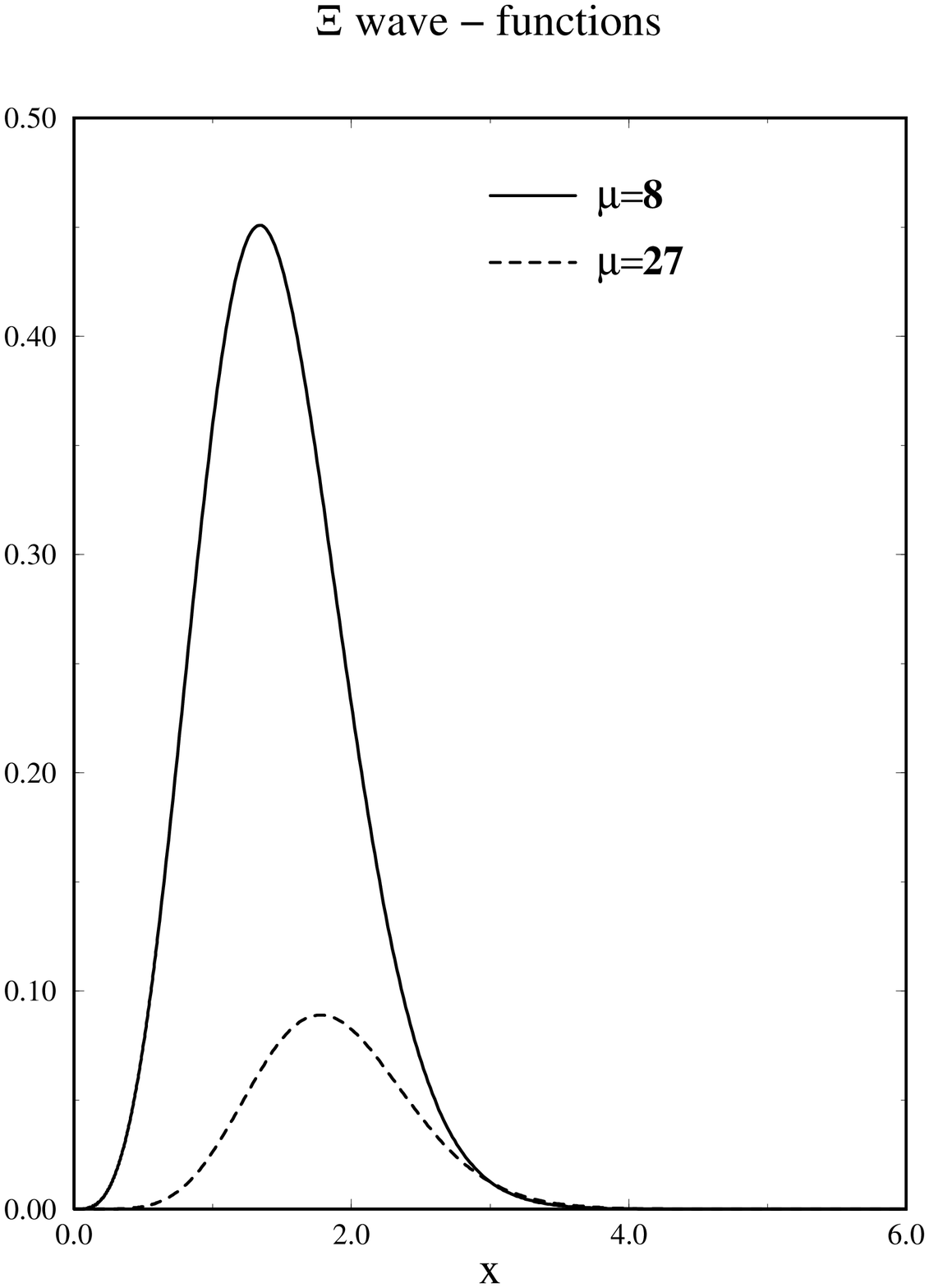,height=6cm,width=7cm}}
\caption{\label{fig_1}
The contributions of the lowest SU(3) representations $\mu$ to the
radial parts of the ground state baryons with $J=1/2$. Here $e=5.0$
is used.}
\end{figure}

In figure \ref{fig_1} the dominant pieces of the radial wave--functions
\be
f^{(B,m)}_\mu(x)=\sum_{n_\mu}C_{\mu,n_\mu}^{(B,m)} f^{(0)}_{\mu,n_\mu}(x)
\label{radwfct}
\ee
are shown for the $\frac{1}{2}^+$ ground states. Here 
$f^{(0)}_{\mu,n_\mu}(x)$ denote the radial eigenfunctions of the 
flavor symmetric formulation (\ref{freebreath}) multiplied 
by $(\alpha^3(x)\beta^4(x)/m(x))^{(1/4)}$\cite{Sch91}.
It should be noted that these radial wave--functions are normalized with 
respect to a metric $m(x)$ which is singular at $x=0$, {\it cf.} eq 
(\ref{defbreath}). Hence all wave--functions vanish at that particular
point. We observe that these ground states are dominated by the radial 
ground state in the octet representation. Nevertheless the contributions 
from the higher dimensional representations are not negligible either. This 
can also be seen from the strange content fractions of these baryons. This 
quantity can be associated with the matrix elements  
$X_S=\langle (1-D_{88})/3\rangle$ \cite{Do86}. 
The sizable deviations from the pure 
octet results are shown in table \ref{tab_2}.
\begin{table}
\caption{\label{tab_2}
The strange content fractions $X_S$ for the $J=1/2$ ground state compared 
to the flavor symmetric case (pure octet). Up to the provided precision the 
results coincide for $e=5.0$ and $e=5.5$. The differences to the pure octet 
case indicate the significance of the higher dimensional representations.}
~
\newline
\centerline{\tenrm\smalllineskip
\begin{tabular}{c | c c c c }
& $N$ & $\Lambda$ & $\Sigma$ & $\Xi$ \\
\hline
$e=5.0$ & $0.16$ & $0.25$ & $0.30$ & $0.37$ \\ 
Octet   & $0.23$ & $0.30$ & $0.37$ & $0.40$ 
\end{tabular}}
\end{table}
It is also interesting to note that the strange content fraction
for the Roper and the $N(1710)$ respectively decrease from 23\% to
14\% and from 25\% to 19\% ($e=5.0$) due to flavor symmetry breaking.

\begin{figure}[t]
~
\vskip1cm
\centerline{\hskip -1.0cm
\epsfig{figure=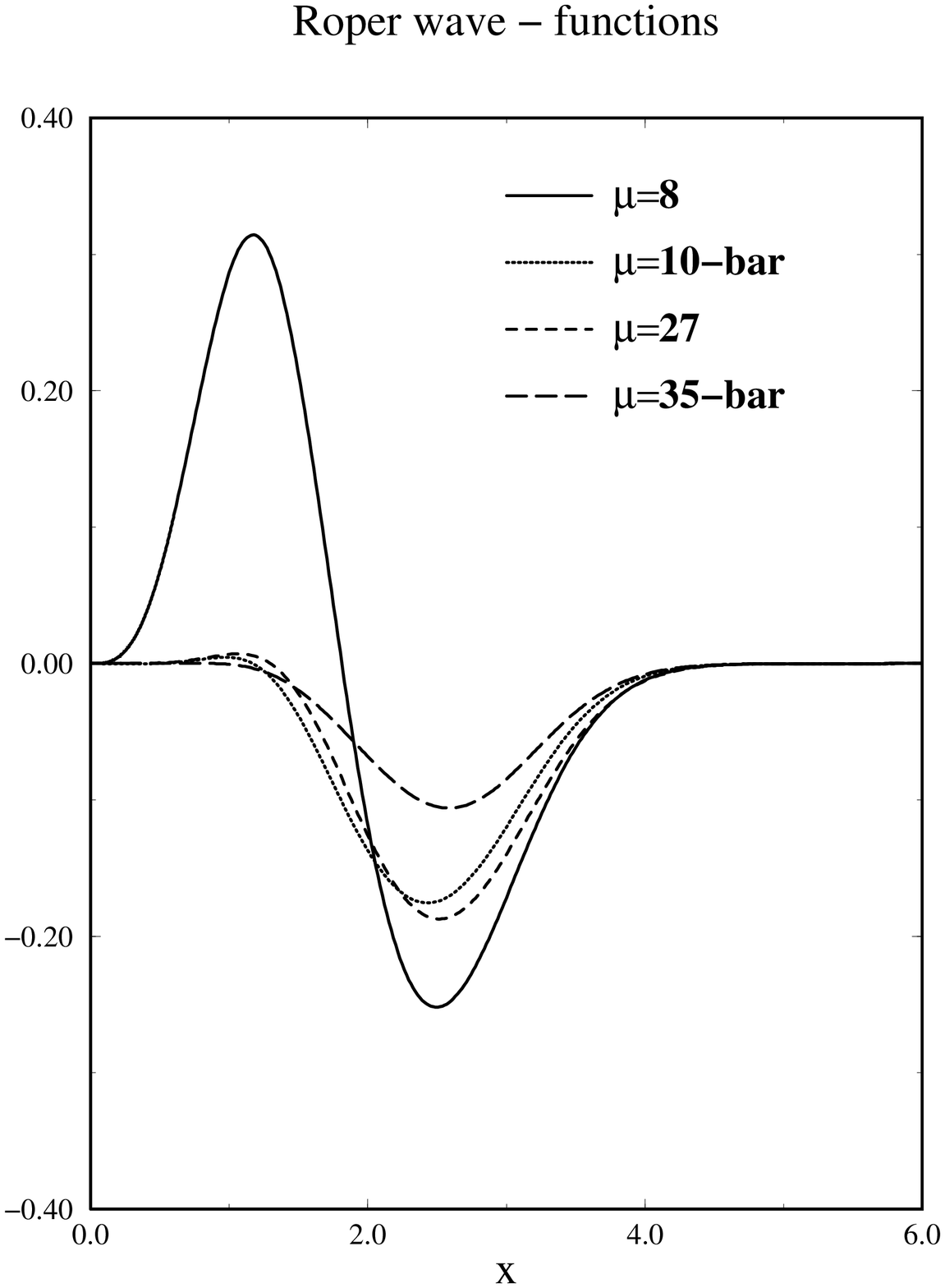,height=7cm,width=7cm}\hskip1cm
\epsfig{figure=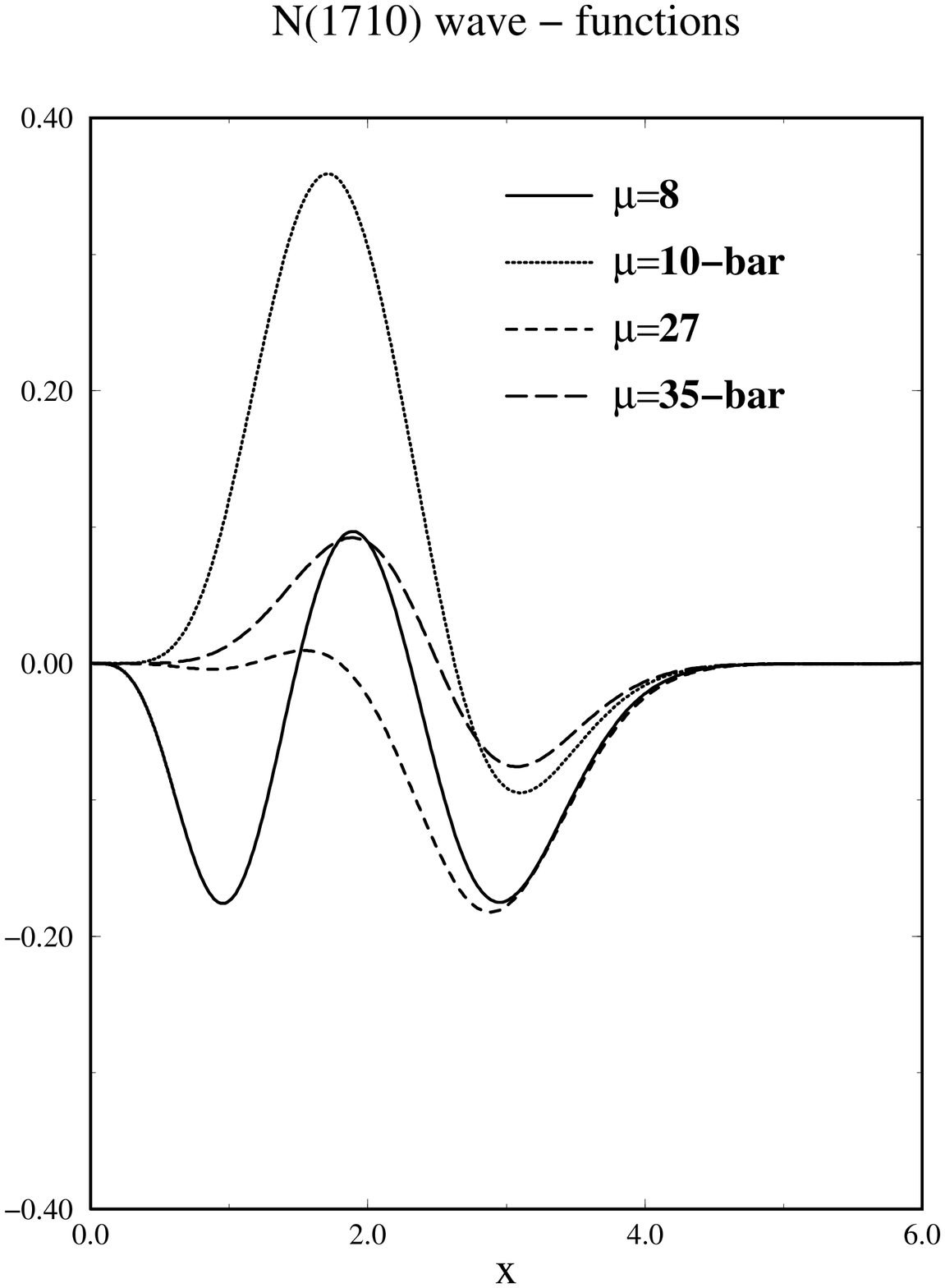,height=7cm,width=7cm}}
\caption{\label{fig_2}
The contributions of the lowest SU(3) representations $\mu$ to the
excited states with nucleon quantum numbers. Here $e=5.0$ is considered.}
\end{figure}

In figure \ref{fig_2} the dependence on the scaling variable for the
first two excited nucleon states is shown. As expected the 
Roper is dominantly a radial excitation of the octet. However,
there are also sizable contributions of the radial ground states
in the higher dimensional representations. 
For the state which we want to associate with the $N(1710)$ we indeed 
find that the $\overline{\mbox{\boldmath $10$}}$ contributes the major 
share. On the other hand the admixture of the $2\hbar\omega$ excitation 
of the octet is not negligible either. Hence the identification 
of the $N(1710)$ with $|N,\overline{\mbox{\boldmath $10$}}\rangle$ 
appears as an over--simplification.

It should be noted that the model predicts yet another eigenstate of 
(\ref{hammatr}) just about $50{\rm MeV}$ above the state we just 
identified as $N(1710)$. This is essentially the linear combination of 
the $|N,\overline{\mbox{\boldmath $10$}}\rangle$ and the $2\hbar\omega$ 
radial excitation of the octet which is orthogonal to the wave--function 
shown in the right panel of figure \ref{fig_2}. Although the Particle 
Data Group (PDG) \cite{PDG} gives $2.1{\rm GeV}$ for the average value of 
the mass of the third resonance in the $P11$ channel there is also an 
analysis \cite{Ma92} of the data which yields a significant lower resonance 
position, $1.885\pm0.030{\rm GeV}$. In particularly one should note that
the four--pole fit of ref \cite{Ba95} predicts two states around 
$1.75{\rm GeV}$ in the $P11$ channel which are less than $10{\rm MeV}$
apart. We find a similar scenario on the $\Sigma$ channel, although about
$200{\rm MeV}$ too high. In table \ref{tab_1} the $|\Sigma, m=2\rangle$ 
state has been considered to be the $\Sigma(1770)\, P11$. In addition we 
observe a $\Sigma$--type state 1.135GeV above the 
nucleon for $e=5.0$ and 1.181GeV for $e=5.5$. Eventually this could be 
identified with the $\Sigma(1880)\, P11$ \cite{PDG}\footnote{In ref 
\cite{Di97} this state was speculated to be the pure 
$|\Sigma,\overline{\mbox{\boldmath $10$}}\rangle$.}. One should bear in 
mind that the analyses leading to this two--star resonance are somewhat 
dated and are spread between\footnote{{\it Cf.} the references compiled 
by the PDG: p. 652 in \cite{PDG}.} $1.826\pm0.020$ \cite{Go80} and 
$1.985\pm0.050{\rm GeV}$ \cite{Va75}.
Nevertheless one is inclined to consider the predicted two almost 
degenerate states in that energy regime as a nice feature of the 
present model. In the $\Lambda$ and $\Xi$ channels no such doubling is 
observed as the ${\overline{\mbox{\boldmath $10$}}}$ does not contain 
states with the quantum numbers of these baryons. However, this 
representation contains a $Y=-1$, $I=\frac{3}{2}$ state which is not
considered here.

States with the quantum numbers of the $Z^+$ exist besides in the 
$\overline{\mbox{\boldmath $10$}}$
also in the $\overline{\mbox{\boldmath $35$}}$, 
$\overline{\mbox{\boldmath $81$}}$ and 
$\overline{\mbox{\boldmath $154$}}$ representations. Actually these
are always the complex conjugates of representations which also 
contain $\Omega$--type states. This is a direct consequence of the
complex conjugation being equivalent to a reflection at the $Y=0$ axis. 
Upon this reflection the $\Xi^*$--type state is transformed into a nucleon 
type state which then satisfies the conditions $J=I$ and $S=0$ while the 
$\Omega$ becomes the $Z^+$. 

\begin{figure}[h]
~
\vskip1.0cm
\centerline{\hskip -1.0cm
\epsfig{figure=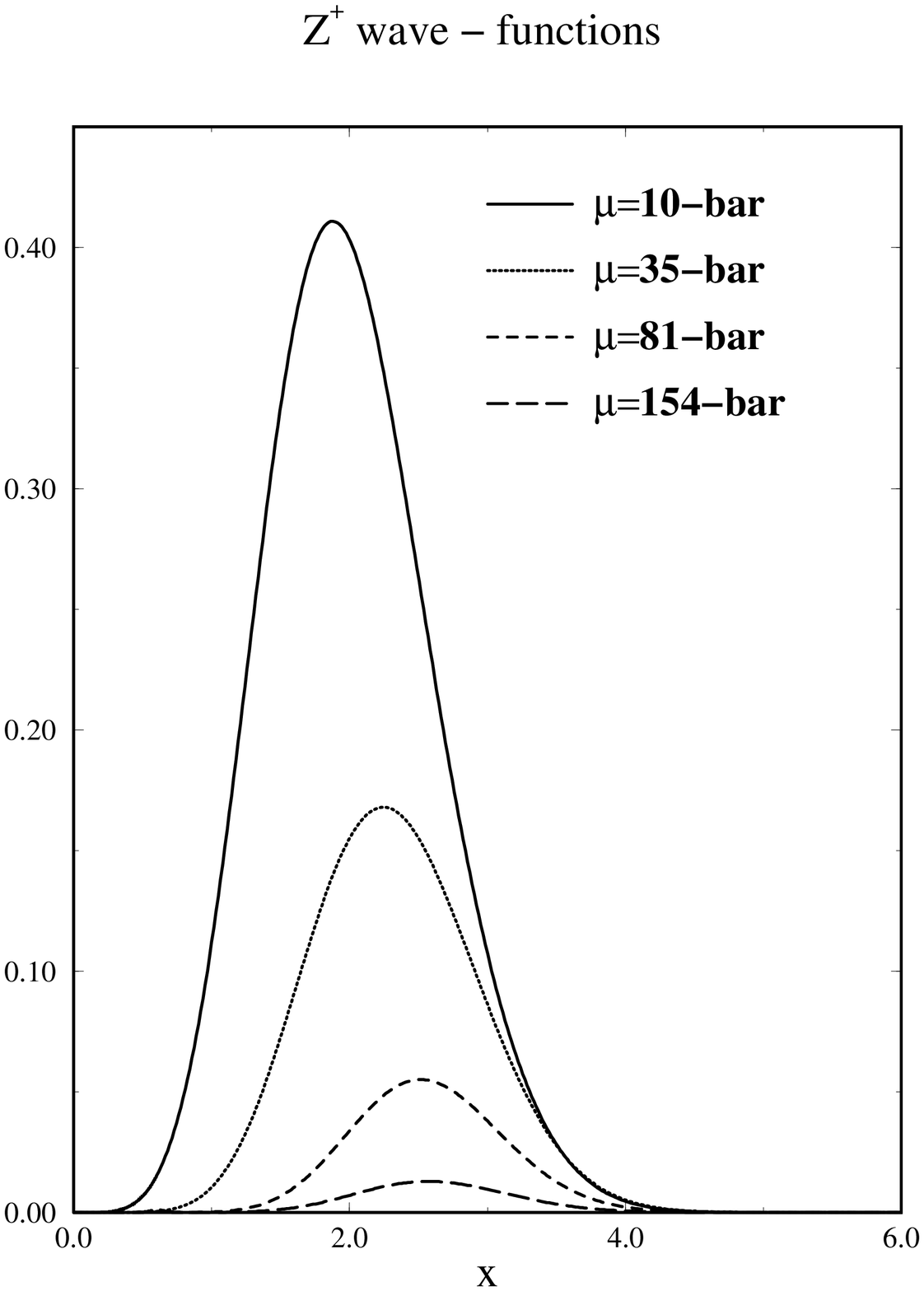,height=7cm,width=7.5cm}\hskip1cm
\epsfig{figure=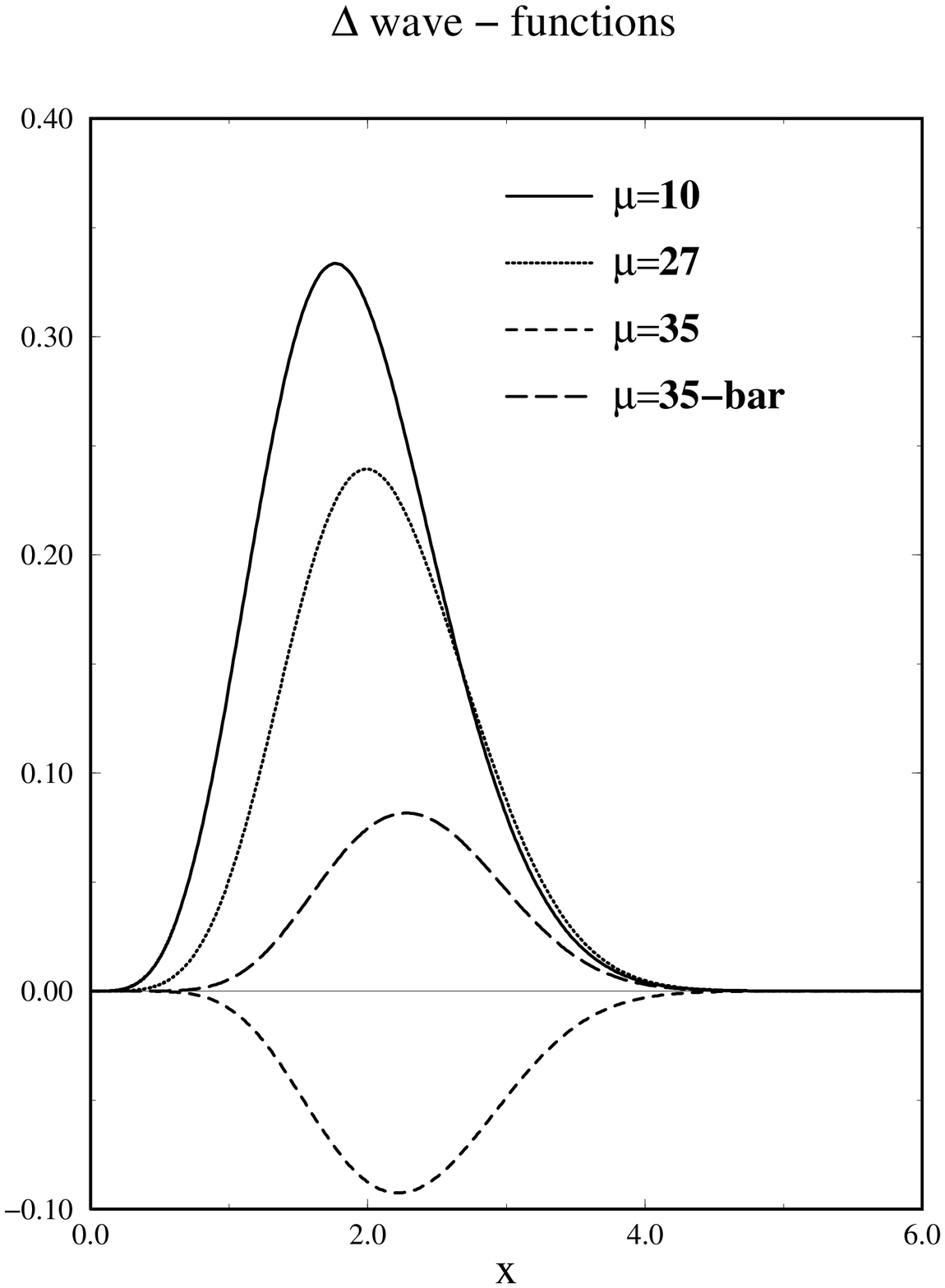,height=7cm,width=7.5cm}}
\caption{\label{fig_3}
The contributions of the lowest SU(3) representations $\mu$ to the 
penta--quark baryon $Z^+$. Also the wave--functions for the 
$\Delta$--resonance are shown. Here $e=5.0$ is considered.}
\end{figure}

The resulting radial structure of the $Z^+$ wave--function is 
displayed in the left panel of  figure \ref{fig_3}.
We recognize that the higher dimensional contributions are not 
negligible. In particular the amplitude of the 
$\overline{\mbox{\boldmath $35$}}$ is almost half as large as the 
leading order piece residing in the $\overline{\mbox{\boldmath $10$}}$. 
It should be noted that up to only first order in flavor symmetry 
breaking the $Z^+$ states in $\overline{\mbox{\boldmath $10$}}$ and 
$\overline{\mbox{\boldmath $35$}}$ have non--vanishing overlap. In the 
estimate of ref \cite{Di97} this overlap was not taken into account. As 
the second order perturbation to the energy of a ground state is always
negative one would speculate that the mass of the $Z^+$ would even be 
reduced when this effect was included. However, here the mass of the 
$Z^+$ is predicted to be $1.57{\rm GeV}$ and $1.59{\rm GeV}$ for $e=5.0$ 
and $e=5.5$, respectively. This is $40$ to $60{\rm MeV}$ larger than 
the result of ref \cite{Di97}. As compared to the octet, the centrifugal 
barrier for states in the $\overline{\mbox{\boldmath $10$}}$ 
representation is stronger. Hence the corresponding eigenfunctions of 
(\ref{freebreath}) are localized at larger values of the scaling variable 
$x$. This effect can be observed by comparing the radial wave--functions
in figures \ref{fig_1} and \ref{fig_3}. In turn it leads to more sizable 
contributions associated with flavor symmetry breaking (\ref{hammatr}) 
than in the model without breathing mode. In order to further illuminate 
the statement that the $Z^+$ wave--function is pushed to larger values 
of $x$ also the wave--function of the $\Delta$ is shown in figure 
\ref{fig_3}. We note that although in leading order the $Z^+$ and 
$\Delta$ have the same Casimir eigenvalue 
$C_2(\overline{\mbox{\boldmath $10$}})=
C_2(\mbox{\boldmath $10$})=6$ the centrifugal barrier is smaller for the 
$\Delta$ because $\alpha(x)>\beta(x)$ in the Hamiltonian (\ref{freebreath}). 
It should be remarked that singular behavior of the metric $m(x)$, which
enters the evaluation of all matrix elements, intensifies this effect.
The increase of the mass due to the $Z^+$ being localized at larger $x$ is 
a leading order effect in flavor symmetry breaking which effects the 
strange content fraction $X_s$. In the flavor symmetric case we find 
$X_s=25\%$ for the $Z^+$. When the flavor symmetry breaking effects are 
included it is reduced to about 18\%. This implies that the $Z^+$ possesses 
a significant cloud of non--strange mesons. It is finally worthwhile to 
note that the first excited state in the $Z^+$ channel is at 
$2.02(2.07){\rm GeV}$ for $e=5.0(5.5)$.

\bigskip
\stepcounter{chapter}
\leftline{\large\it 3. Estimate of Widths}
\smallskip

The soliton model described in the preceding section has been shown to 
reasonably describe not only the spectrum of the low--lying $\frac{1}{2}^+$ 
and $\frac{3}{2}^+$ baryons but also various baryon static properties. In 
particular the inclusion of the radial collective coordinate properly
reproduces the experimentally observed deviation from U--spin symmetry of 
the predictions for the baryon magnetic moments \cite{Sch91,Sch91a}. This 
deviation remains unobserved as long as flavor symmetry breaking effects 
on the extension of the soliton are not included \cite{Schw92,We96}. 
Hence one is inclined to assume that also the widths for various decays of 
the predicted states (resonances) are reasonably described. Here we are 
interested in the decay $Z^+\to K^{(+,0)} N$ which should be mediated by 
a pseudoscalar Yukawa coupling. In soliton models such a coupling is not 
directly obtainable as to leading order in $1/N_C$ terms linear in the meson 
fluctuations vanish by definition. One possibility to avoid this problem 
is to adopt the Goldberger--Treiman relation, which relates the relevant
coupling constant to the axial charge. Along that line the matrix element 
of $D_{33}$ was used in ref \cite{Ad83} to predict the amplitude for the 
decay $\Delta\to N\pi$ and relate it to the $\pi N$ coupling constant. In 
the present model the situation is even less transparent as we also demand 
the dependence of the transition operator on the scaling variable $x$. 
Assuming, for the time being, that we have obtained the relevant operator 
in the space of the collective coordinates the corresponding matrix element 
will yield the coupling constant $G_{B^\prime\to B \phi}$ associated with 
the decay of the resonance $B^\prime$ to another baryon state $B$ and a 
meson $\phi$. This coupling constant then enters the width via
\be
\Gamma(B^\prime\to B \phi)=\frac{3G_{B^\prime\to B \phi}^2}
{8\pi M_{B^\prime}M_B} \ |\mbox{\boldmath $p$}_\phi|^3 \ .
\label{width1}
\ee
Here $|\mbox{\boldmath $p$}_\phi|$ is the momentum of the outgoing meson 
in the rest frame of the resonance $B^\prime$. The cubic dependence on 
$|\mbox{\boldmath $p$}_\phi|$ arises as the amplitude (which enters the 
width quadratically) of the pseudoscalar Yukawa coupling is linear in 
$|\mbox{\boldmath $p$}_\phi|$. In addition the phase--space provides one
power in that momentum. For states which are located just slightly above
threshold this cubic dependence on the momentum rather than the value of 
the coupling constant will be the most crucial ingredient to calculate the 
width of $B^\prime$. Hence it is sufficient to get a rough estimate for 
the coupling constants in order to allow for a comparison of various decay 
processes. It is therefore suggestive to adopt the following 
strategy: For the flavor part of the relevant operator we adopt 
$D_{\phi,3}$ which in leading order $1/N_C$ is the only possible operator 
compatible with flavor covariance. For the scaling piece we will consider
different powers, $\mu^{-n}=x^{2n/3}$. Different values for the power 
$n$ can be motivated by the long range behavior of the pseudoscalar fields 
which built up the soliton. Taking straightforwardly the matrix element 
of the pion field would result in $n=3$ from the spatial integration. 
Considering that due to the pseudoscalar character of that field the 
Fourier transformation involves the spherical Bessel function $j_1(qr)$ 
would add a factor $r$ to the integrand \cite{Me89}, whence $n=4$. On the 
other hand one could argue that in the chiral limit ($m_\pi=0$) the coupling 
constant is directly related to the amplitude of the soliton at large $r$ 
\cite{Do94}. As the massless pion field decays like $1/r^2$ one would be 
inclined to adopt $n=2$. In this way we will obtain at least the 
generic behavior of the coupling constant while the major ingredient for 
the decay width, the momentum $|\mbox{\boldmath $p$}_\phi|$, is computed 
from the spectrum calculated in the preceding section. The widths of 
various decays will finally be compared by adjusting the absolute magnitude 
to the process $\Delta\to N \pi$, {\it i.e.}
$\Gamma(\Delta\to N \pi)\approx 120{\rm MeV}$. 
This corresponds to a multiplicative normalization of the decay 
constants which is also suggested by large--$N_C$ considerations
\cite{Da94}.

To be precise, we will compute matrix elements of the form
\be
G_{B^\prime\to B \phi}= C_{\Delta}
\langle B^\prime m^\prime | x^{2n/3} D_{a3} | B m \rangle ,
\label{width2}
\ee
with $C_{\Delta}$ fitted to $\Gamma(\Delta\to N \pi)$. Here we take $a=3$ 
and $a=4\pm i5$ for strangeness conserving ($\phi=\pi$) and strangeness
changing ($\phi=K$) decays, respectively. The latter case is only relevant 
for the decay of the $Z^+$. The baryon wave--functions are those of 
eq (\ref{bsbreath}) which stem from diagonalizing the full collective 
Hamiltonian. 
\begin{table}
\caption{\label{tab_3}Decay widths (in MeV) and ratio of $\pi N$ and 
$\pi\Delta$ coupling constants using the matrix elements 
(\protect\ref{width2}). The decay width 
$\Gamma(\Delta\to N \pi)\approx 120{\rm MeV}$ is fixed. $R$ denotes the 
Roper (1440) resonance. Experimental data are extracted 
from ref \protect\cite{PDG}.}
~
\newline
\centerline{\small\smalllineskip
\begin{tabular}{c | c c c | c c c| c}
& \multicolumn{3}{c|}{e=5.0} & \multicolumn{3}{c|}{e=5.5} & expt. \\
\hline
$n$ & 4 & 3 & 2 & 4 & 3 & 2 & \\
\hline
$\Sigma^*\to\Sigma\pi$  & 1 & 1 & 1 & 2 & 2 & 2 & $4\pm1$ \\
$\Sigma^*\to\Lambda\pi$ & 33& 38& 42& 37& 38& 43& $32\pm4$ \\
$\Xi^* \to \Xi \pi $    & 5 & 7 & 10& 7 & 9 & 11& $10\pm2$ \\
\hline
$R\to N\pi $ & 429 & 281 & 156 & 424 & 260 & 145 & 200 to 320 \\
$R\to\Delta\pi $ & 4 & 2 & 2 & 9 & 6 & 3 & 50 to 80 \\
\hline
$Z^+\to N K$ & 118 & 121 & 124 & 130 & 124 & 126 & ? \\
\hline
$g_{\pi NN}/g_{\pi N\Delta} $ &
0.77 & 0.79 & 0.83 & 0.77 & 0.79 & 0.83 & 0.68 
\end{tabular}}
\end{table}
The resulting widths as well as the ratio of the coupling constants
between the $\Delta$ and the nucleon to the pion are shown in 
table \ref{tab_3}. Let us recall that as $\Gamma(\Delta\to N \pi)$ is 
kept fixed one should consider $g_{\pi N\Delta}$ as an input quantity.

Apparently we find that, at least for the widths of ground state
baryons, the dependence on the power $n$ is only moderate. The reason
is that for these baryons the major contribution to the scaling part of 
the matrix element stems from the vicinity of $x=1$. In addition the
shape of the wave--functions of these baryons is quite similar in that
region, {\it cf.} figure \ref{fig_1}. Hence the effect of different $n$ 
is compensated by normalizing to $\Gamma(\Delta\to N \pi)$. Regarding 
the crudeness of our estimate the predicted widths for the processes 
$\Sigma^*\to\Sigma\pi$, $\Sigma^*\to\Lambda\pi$ and $\Xi^*\to\Xi\pi$
as well as the ratio $g_{\pi NN}/g_{\pi N\Delta} $ are in fair 
agreement with the empirical data. The case of the Roper (1440) 
resonance is different. Here we recognize a strong dependence on 
the power $n$. This is a consequence of the associated breathing mode
wave--function having a node around $x=1.5-2.0$, {\it cf.} figure 
\ref{fig_2}. For the decay $R\to N\pi$ 
the value $n=3$ appears to be reasonable. However, one should be careful 
with such a conclusion as the too low prediction for the mass of the 
Roper might falsify the result for the width. This is even more 
pronounced for the process $R\to\Delta\pi$. Table \ref{tab_3} indicates 
that at least one order of magnitude is missing for the width. For the 
masses given in table \ref{tab_1} the momentum of the outgoing pion is 
66 (91)MeV for $e=5.0(5.5)$. Substituting the physical momentum, 
$|\mbox{\boldmath $p$}_\pi|=147{\rm MeV}$ could account for an order of 
magnitude for the decay width (\ref{width2}). Again we recognize that the 
decay widths are significantly more sensible to the mass parameters than 
to the decay constants, in particular for processes with kinematics just 
above threshold. For the decay of the $Z^+$, the process we are mostly 
interested in, we recognize neither a strong dependence on the power $n$ 
nor on the model parameter $e$. From the results shown in table \ref{tab_3} 
it seems fair to conclude that the width of the process $Z^+\to N K$ follows 
closely the width of $\Delta\to N \pi$.

Rather than just identifying the matrix element of the pseudoscalar 
field with the coupling constant $G_{B^\prime\to B \phi}$ one could 
imagine to compute this coupling constant via the axial current 
and adopt the Goldberger--Treiman relation. In three flavor space
this is different from the above approach because an additional
operator, whose contribution to $G_{B^\prime\to B \phi}$ is suppressed 
by $1/N_C$, enters the calculation. In this approach one calculates 
the matrix elements
\be
G_{B^\prime\to B \phi}\sim
\langle B^\prime m^\prime | \left\{
\left(g_1 x^{4/3} +g_2 \right) D_{a3} 
+g_3 \frac{x^{2/3}}{\beta(x)}
\sum_{\alpha,\beta=4}^7 d_{3\alpha\beta}D_{a\alpha}R_\beta
\right\}| B m \rangle \ .
\label{width3}
\ee
The constants $g_1,g_2$ and $g_3$ are functionals of the static 
soliton and can be extracted from refs \cite{Sch91,Sch91a}. The additional
operator involves the right SU(3)--generators $R_a$. As it is multiplied 
by the inverse moment of inertia for rotations into strange direction
the contribution of this operator to the coupling constant will be 
suppressed by $1/N_C$. Hence the adjustment of the coupling constants
to the decay $\Delta\to N\pi$ in the spirit of the large--$N_C$ 
expansion \cite{Da94} requires to only normalize $g_1$ and $g_2$
rather than the whole matrix element (\ref{width3}). The numerical
results for that calculation are shown in table \ref{tab_4}.
\begin{table}
\caption{\label{tab_4}Decay widths (in MeV) and ratio of $\pi N$ and 
$\pi\Delta$ coupling constants using the matrix elements 
(\protect\ref{width3}). The decay width 
$\Gamma(\Delta\to N \pi)\approx 120{\rm MeV}$ is fixed. $R$ denotes the 
Roper (1440) resonance. Experimental data are extracted 
from ref \protect\cite{PDG}.}
~
\newline
\centerline{\small\smalllineskip
\begin{tabular}{c | c | c | c}
& e=5.0 & e=5.5 & expt. \\
\hline
$\Sigma^*\to\Sigma\pi$  & 2  & 3 & $4\pm1$ \\
$\Sigma^*\to\Lambda\pi$ & 64 & 63 & $32\pm4$ \\
$\Xi^* \to \Xi \pi $    & 22 & 23 & $10\pm2$ \\
\hline
$R\to N\pi $ & 71 & 71 & 200 to 320 \\
$R\to\Delta\pi $ & 2 & 3 & 50 to 80 \\
\hline
$Z^+\to N K$ & 82 & 83 & ? \\
\hline
$g_{\pi NN}/g_{\pi N\Delta} $ &
0.70 & 0.69 & 0.68 
\end{tabular}}
\end{table}
The operator (\ref{width3}) does not seem to be very well suited in 
particular because the width for the Roper decaying into a nucleon and a 
pion is significantly underestimated. We note that the dependence 
on the momentum of the outgoing meson cannot be made responsible 
for the short--coming in this process as the resonance is far away from 
threshold. For $e=5.5$ we have $|\mbox{\boldmath $p$}_\pi|=354{\rm MeV}$ 
which is not too different from the physical value of 396MeV. Hence using 
the physical masses would at best give a 40\% increase of the width. Also 
the widths for the decays $\Sigma^*\to\Lambda\pi$ and $\Xi^* \to\Xi\pi$ 
turn out to be somewhat too large. On the other hand the width for $Z^+$ 
is slightly reduced as compared to the use of eq (\ref{width2}). This is 
mainly due to the fact that the two SU(3) operators in (\ref{width3}) 
interfere destructively for the state 
$|Z^+,\overline{\mbox{\boldmath $10$}}\rangle$; 
contrarily they interfere constructively for the ordinary baryons. 
To be precise, $\langle \Delta, {\mbox{\boldmath $10$}} | D_{33} 
| N, {\mbox{\boldmath $8$}} \rangle=(1/2)
\langle \Delta, {\mbox{\boldmath $10$}} |
\sum_{\alpha,\beta=4}^7 d_{3\alpha\beta}D_{3\alpha}R_\beta
| N, {\mbox{\boldmath $8$}} \rangle$ while
$\langle Z^+, {\overline{\mbox{\boldmath $10$}}} | D_{K3}
| N, {\mbox{\boldmath $8$}} \rangle=-(1/2)
\langle Z^+, {\overline{\mbox{\boldmath $10$}}} |
\sum_{\alpha,\beta=4}^7 d_{3\alpha\beta}D_{K\alpha}R_\beta
| N, {\mbox{\boldmath $8$}} \rangle$. 

To summarize this section it seems reasonable to state that the breathing 
mode approach to the $Z^+$ predicts a width of that state of the order of 
100MeV. This is considerably larger than the prediction of 15MeV found in ref 
\cite{Di97}. As discussed intensively, a major reason for this 
difference is not the difference in the coupling constant for the process 
$Z^+\to K N$ but rather the larger mass found for the $Z^+$ in the 
present approach. As compared to ref \cite{Di97} the result for the mass 
of the $Z^+$ has increased only moderately by $50{\rm MeV}$. Nevertheless 
it has noticeable consequences for the width of the only possible decay 
mode of this penta--quark state, $Z^+\to N K$. The momentum of the outgoing 
kaon grows from $|\mbox{\boldmath $p$}_K|=254{\rm MeV}$ to 320MeV 
increasing the width by a factor of two as for processes which are just 
above threshold the momentum of the outgoing meson is a quickly rising 
function of the resonance position. In order to further compare the 
width of the $Z^+$ with the result of ref \cite{Di97} it should be noted 
that such a comparison should concern the $80{\rm MeV}$ displayed in 
table \ref{tab_4} because those authors also included the 
$\sum_{\alpha,\beta=4}^7 d_{3\alpha\beta}D_{a\alpha}R_\beta$ operator. 
As discussed this operator lowers the prediction for the width of the 
$Z^+$ due to the destructive interference with the leading operator 
$D_{a3}$. The moment of inertia for rotations into strange directions 
($\beta(x)$) appears in the denominator of the additional operator. 
Hence the contribution of this operator will be most sensible to the 
small--$x$ shape of the wave--function. As already discussed, the 
wave--function for the $Z^+$ penta--quark is more pronounced at larger
values of $x$ due to the angular barrier being stronger for states in 
the higher dimensional SU(3) representations. As indicated in figure 
\ref{fig_3} this is also the case when we compare with the $\Delta$ 
wave--function whose matrix elements set the scale for our estimate of 
the width. As a result the contribution of the additional SU(3) 
operator is reduced even further. However, this is only a 20--30\% 
effect and still does not explain the full discrepancy with ref 
\cite{Di97}. At this point one should note that in ref \cite{Di97} the 
numerical results for the widths of the $\frac{3}{2}^+$ baryons are 
erroneous\footnote{For example, for the process $\Delta\to N\pi$ the use 
of eq (\cite{Di97}:42) together with the empirical values for 
the masses of the involved hadrons and the suggested coupling constant
$G_0=19$ yields a width of 64MeV rather than the alleged 110MeV. 
However, the expression (\cite{Di97}:56) for the width of the $Z^+$ has 
been worked out correctly. As an attempt to locate the possible error it 
could be remarked that the replacement of the factor $M_2/M_1$ by its 
inverse in eq (\cite{Di97}:49) yields the numerical results presented 
in eqs (\cite{Di97}:42)--(\cite{Di97}:45) for the decay widths of the 
${3/2}^+$ baryons. The analogous replacement in eq (\cite{Di97}:56) 
results in a $Z^+$ width of about 40MeV \cite{Po98}.}. As those 
overestimated widths have subsequently been employed to set the overall 
scale this is likely to be the reason for the remaining discrepancy.

\bigskip
\stepcounter{chapter}
\leftline{\large\it 4. Conclusions}
\smallskip

In the present study we have investigated the coupling between radial 
and (flavor) rotational motion of a chiral soliton in flavor SU(3). 
Upon canonical quantization of the corresponding collective 
coordinates this approach not only describes the spectrum of the
low--lying $\frac{1}{2}^+$ and $\frac{3}{2}^+$ baryons but also that
of the excited states in the respective channels. Besides mixing of 
various SU(3) representations the model in particular may account for 
an eventual resonance doubling \cite{Ba95} in the nucleon P11 channel 
around 1.75GeV. A similar scenario is observed for the $\Sigma$ channel. 
These results provide additional support for identifying baryon states 
in the ${\overline{\mbox{\boldmath $10$}}}$ representation of flavor 
SU(3) with observed resonances. Subsequently this picture leads to the 
question of properly identifying those baryon states in such higher 
dimensional representations which do not have counterparts in the octet or 
decuplet. In this respect the $Z^+$ ($Y=2$, $I=0$, $J^\pi=\frac{1}{2}^+$) 
is the most interesting candidate as probably being the lightest one. 
Previously \cite{Di97} this state was considered to be a pure 
${\overline{\mbox{\boldmath $10$}}}$ baryon. That calculation, however, 
was not a full model calculation but rather a compilation of possible 
terms allowed by the flavor symmetries of the model. The constants 
of proportionality were determined from the known baryon spectrum and 
radial degrees of freedom were frozen. In this treatment it seems 
doubtful to {\it ad hoc} identify of the $N(1710)$ (potentially 
a radially excited nucleon) with the nucleon state in the 
${\overline{\mbox{\boldmath $10$}}}$ representation. Here we have 
reflected on that assumption by carrying out a full model calculation 
and emphasizing on the admixture of both, higher dimensional SU(3) 
representation as well as radially excited baryon states which otherwise 
have identical quantum numbers. It should be noted that both types of
admixture are mediated through flavor symmetry breaking. Despite these 
major extensions of the model treatment, the present prediction for the  
mass of the $Z^+$, 1.58GeV, is only about 50MeV higher than that of ref 
\cite{Di97} but still about $60{\rm MeV}$ lower than the value found in 
ref \cite{Wa92}. As discussed, in the soliton model the determination of 
the coupling constants for various decays bears quite some uncertainties, 
nevertheless the model calculations suggest 
$\Gamma(Z^+\to N K)\sim 100{\rm MeV}$ as an estimate for the width of 
the $Z^+$. Quite a substantial uncertainty should be attributed to this 
value. Comparison of the different estimates collected in section 3 
suggests $\pm30{\rm MeV}$. This should be considered a lower bound 
for the uncertainty.

Here we have employed a soliton model which besides the pseudoscalar octet 
mesons contains a scalar field. This scalar field has been introduced as to 
mock up the QCD trace anomaly. Although this is presumably not the most 
natural choice for an effective meson theory, we have motivated this 
model from the simplicity to include the breathing degree of freedom
and its previous success to reasonably describe the spectrum of the 
low--lying $\frac{1}{2}^+$ and $\frac{3}{2}^+$ baryons as well as 
various of their static properties.

Finally one could object that the prediction of {\it exotic} states like 
$Z^+$ would completely be due to the adopted quantization scheme for 
the flavor degrees of freedom. In the alternative bound state approach
\cite{Ca85} a penta--quark state with the quantum numbers of  the $Z^+$ 
would emerge as a bound system of the soliton and a kaon, while the 
ordinary hyperons are considered as anti--kaons bound in the soliton 
background. Such penta--quark states are found to be unbound unless the 
kaon mass is artificially tuned to about $1{\rm GeV}$. However, the 
resonance doubling found in the nucleon and $\Sigma$ channels around 
$2{\rm GeV}$ is not without experimental support which indicates that
{\it exotic} representations like the ${\overline{\mbox{\boldmath $10$}}}$
indeed have physical significance.

\bigskip
\leftline{\large\it Acknowledgments}
\smallskip
I am grateful for valuable discussions with R. Alkofer, 
H. Reinhardt, J. Schechter and H. Walliser. W. Eyrich is thanked for 
pointing out the relevance of the present study while visiting the
T\"ubingen Graduiertenkolleg {\it Hadronen und Kerne}. Furthermore I would
like to thank M. Polyakov for bringing ref \cite{Ba95} to my attention 
and for discussions on details of ref \cite{Di97}. 

This work is supported by the Deutsche 
Forschungsgemeinschaft (DFG) under contract Re 856/2-3.

\end{document}